\documentclass[12pt,a4paper,reqno]{amsart}
\usepackage[T1]{fontenc}
\usepackage{bbm}
\usepackage{graphicx}
\usepackage{blkarray}
\usepackage{tikz}
\usepackage{amsmath}
\usepackage{amssymb}
\usepackage{amssymb}
\usepackage{bbold}
\usepackage{mathrsfs}
\usepackage{amsthm}
\usepackage{amsfonts}
\usepackage{color}
\usepackage{tikz}
\usepackage{hyperref}
\usetikzlibrary{calc}
\usepackage{float}
\usepackage[shortlabels]{enumitem}
\theoremstyle{plain} 

\newtheorem{proposition}
{Proposition}[section]

\newtheorem{Conjecture}
{Conjecture}[section]
\newtheorem{property}
{Property}[section]
\theoremstyle{definition} 
\newtheorem{definition}{Definition}[section]
\newtheorem{example}{Example}[section]
\theoremstyle{remark} 

\begin{document}
\title{Quantum Stabilizer formalism for any composite system}
\author{Zhelin Tian \vspace{0.3cm} \\  \footnotesize University of York, UK}

\begin{abstract}
    In this paper, we try to generalise quantum stabilizer formalism\cite{Daniel}\cite{Chuang}\cite{wY} to any composite system, that is, it includes not only composite systems of equal dimensions, but also composite systems of unequal dimensions.
\end{abstract}
\maketitle
\tableofcontents
\section{Definition of Qudit}

\begin{definition}
    \text{Qudit} is a vector in $d$-dimensional Hilbert space, satisfying
    \begin{equation*}
        |\psi\rangle = \sum_{i=0}^d \alpha_i |i\rangle 
    \end{equation*}
Where $\sum_{i=1}^d |\alpha_i|^2 = 1$, $d \in \mathbb{N}$ and $\{|i\rangle,i\in \mathbb{Z}_d\}$ are orthonomal basis.
\end{definition}
\begin{example}
    For $d=2$,
\begin{equation*}
    |\psi\rangle = \alpha_0|0\rangle + \alpha_1|1\rangle
\end{equation*}
where $\sum_{i=0}^{1}|\alpha_i|^2=1$. They are called qubit in the literature.
\end{example}
\section{Extended Pauli operators and Weyl operators}

\begin{definition}
    In a $d$-dimensional Hilbert space, define $X_d$:
    \begin{equation*}
        X_d|x\rangle = |x+1\rangle\mod{d}
    \end{equation*}
And define $Z_d$:
\begin{equation*}
    Z_d|x\rangle = \omega^x|x\rangle
\end{equation*}
Where $\omega = e^{\frac{2\pi i }{d}}$, $x\in\mathbb{Z}_d$. they are called \textbf{Extended Pauli operators}
\end{definition}
Remarks:
\begin{itemize}
    \item Pauli matrices $I, X, Y, Z$ defined as:
    \begin{equation*}
    \sigma_0 \equiv I = \begin{bmatrix}
        1 & 0\\
        0 & 1\\
    \end{bmatrix},\sigma_1 \equiv X= \begin{bmatrix}
        0 & 1\\
        1 & 0\\
    \end{bmatrix}
\end{equation*}
\begin{equation*}
    \sigma_2 \equiv Y= \begin{bmatrix}
        0 & -i\\
        i & 0\\
    \end{bmatrix},\sigma_3 \equiv Z= \begin{bmatrix}
        1 & 0\\
        0 & -1\\
    \end{bmatrix}
\end{equation*}
In our notation, $X \equiv X_2$ and $Z \equiv Z_2$.
\end{itemize}
\begin{property}
       $\omega X_dZ_d = Z_dX_d$ 
\end{property}

\begin{definition}
    Define operators $W_{p,q}$
    \begin{equation*}
        W_{p,q} = X_d^pZ_d^q
    \end{equation*}
where $p,q\in \mathbb{Z}_d$, they are called \textbf{Weyl operators}.
\end{definition}
\begin{property}
    Suppose $W_{p,q} = X_d^pZ_d^q$ and $W_{p',q'} = X_d^{p'}Z_d^{q'}$, then
    \begin{equation*}
        W_{p,q}W_{p',q'} = \omega^{qp'} W_{p+p',q+q'}
    \end{equation*}
\end{property}
\section{Global Pauli Group}

\begin{definition}
    Define \textbf{Global Pauli Group} -  
    $G_{k_1\times k_2\cdots\times k_n} = \{(e^{\frac{i\pi}{\text{lcm}\{k_1,k_2,\cdots,k_n\}}})^c X_{k_1}^{p_1}Z_{k_1}^{q_1}\otimes X_{k_2}^{p_2}Z_{k_2}^{q_2}\otimes\cdots\otimes X_{k_n}^{p_n}Z_{k_n}^{q_n},k_i\in \mathbb{N},q_i,p_i\in \mathbb{Z}_{k_i},c\in\mathbb{Z}_{2\times\text{lcm}\{k_1,k_2,\cdots,k_n\}}\}$, where $\text{lcm}\{k_1,k_2,\cdots,k_n\}$ is least common multiple of $k_1,k_2,\cdots,k_n$.
\end{definition}
\begin{proposition}
    $G_{k_1\times k_2\cdots\times k_n}$ is well-defined, that is, it satisfies the definition of a group.
\end{proposition}
\begin{proof}
    Suppose $g$ and $g'$ are
    \begin{equation}
        g = (e^{\frac{i\pi}{\text{lcm}\{k_1,k_2,\cdots,k_n\}}})^{c}X_{k_1}^{p_1}Z_{k_1}^{q_1}\otimes\cdots\otimes X_{k_n}^{p_n}Z_{k_n}^{q_n}
    \end{equation}
    \begin{equation}
        g' = (e^{\frac{i\pi}{\text{lcm}\{k_1,k_2,\cdots,k_n\}}})^{c'}X_{k_1}^{p'_1}Z_{k_1}^{q'_1}\otimes\cdots\otimes X_{k_n}^{p'_n}Z_{k_n}^{q'_n}.
    \end{equation}
    \begin{equation}
        gg' = (e^{\frac{i\pi}{\text{lcm}\{k_1,k_2,\cdots,k_n\}}})^{c+c'}(e^{\frac{2\pi i}{k_1}})^{q_1p'_1}\cdots(e^{\frac{2\pi i}{k_n}})^{q_np'_n}X_{k_1}^{p_1+p'_1}Z_{k_1}^{q_1+q'_1}\otimes\cdots\otimes X_{k_n}^{p_n+p'_n}Z_{k_n}^{q_n+q'_n}
    \end{equation}
Because $\text{lcm}\{k_1,k_2,\cdots,k_n\} = k_ih_i$, where $h_i\in\mathbb{N}$, so
\begin{equation}
    gg' = C X_{k_1}^{p_1+p'_1}Z_{k_1}^{q_1+q'_1}\otimes\cdots\otimes X_{k_n}^{p_n+p'_n}Z_{k_n}^{q_n+q'_n}
\end{equation}
where $C$ is
\begin{equation}
    C= (e^{\frac{i\pi}{\text{lcm}\{k_1,k_2,\cdots,k_n\}}})^{c+c'+2h_1q_1p'_1+\cdots+2h_nq_np'_n}
\end{equation}
Due to 
\begin{equation}
(c+c'+2h_1q_1p'_1+\cdots+2h_nq_np'_n) \mod\{\text{lcm}\{k_1,k_2,\cdots,k_n\}\} \in\mathbb{Z}_{2\times\text{lcm}\{k_1,k_2,\cdots,k_n\}}
\end{equation}
So it can be proved that $G_{k_1\times k_2\cdots\times k_n}$ is closed. The other axioms in the definition of a group are not difficult to verify.
\end{proof}

\begin{property}
    $|G_{k_1\times k_2\cdots\times k_n}| = 2\times\text{lcm}\{k_1,k_2,\cdots k_n\}\times(k_1k_2\cdots k_n)^2$
\end{property}
\begin{example}
    When $k_1 = k_2 = \cdots = k_n = 2$, we can see that,
\begin{equation}
    |G_{2\times 2\cdots\times 2}|=|G_{2^n}|= 2\times2\times(2^n)^2 = 4\times4^n = 4^{n+1}
\end{equation}
\end{example}
\section{Definition of Stabilizer}
\begin{definition}
    If $S$ is a subgroup of $G_{k_1\times k_2\cdots\times k_n}$, and $V_{k_1\times k_2\cdots\times k_n}= \mathbb{C}^{k_1}\otimes\mathbb{C}^{k_2}\otimes\cdots\otimes\mathbb{C}^{k_2}$, $|\psi\rangle\in \mathbb{C}^{k_1}\otimes\mathbb{C}^{k_2}\otimes\cdots\otimes\mathbb{C}^{k_n}$, $V_S$ is defined as
    \begin{equation}
        V_S \equiv \{|\psi\rangle ~|~g|\psi\rangle = |\psi\rangle,\forall g\in S,\forall |\psi\rangle \in V_{k_1\times k_2\cdots\times k_n}\}
    \end{equation}
    
    $V_S$ is called \textbf{stabilized} by  subgroup $S$, $S$ is called \textbf{stabilizer} of $V_S$.\\
 \end{definition}   
We can also define $V_S$ as the intersection of the eigenvalue one eigenspace of each element of stabilizer $S$. The following proposition gives a detailed explanation.
\begin{proposition}
    If $S = \{g_1,g_2,\cdots,g_l\}$ is stabilizer of $V_S$, then $V_S$ is equal to the intersection of  $V_{g_i}$ for every $g_i \in S $, where $i\in\{1,2,\cdots,l\}$.
    \begin{equation}
         V_S \equiv \{|\psi\rangle | g_i|\psi\rangle = |\psi\rangle,\forall g_i\in S,\forall |\psi\rangle \in V_n\} = \bigcap_{g_i\in S} V_{g_i}
    \end{equation}\label{eq:34}
\end{proposition}
\begin{proof}
    First part:
    \begin{equation}
        |\psi\rangle \in \bigcap_{g_i\in S} V_{g_i} \implies \forall g_i\in S,|\psi\rangle \in V_{g_i} \implies \forall g_i\in S,g_i|\psi\rangle=|\psi\rangle\implies|\psi\rangle\in V_S
    \end{equation}
So
\begin{equation}
    \bigcap_{g_i\in S}V_{g_i} \subset V_S
\end{equation}
Second part:
\begin{equation}
\begin{split}
     |\psi\rangle\in V_S &\implies \forall g_i\in S,g_i|\psi\rangle=|\psi\rangle \implies \forall g_i\in S,|\psi\rangle\in V_{g_i}\\
     &\implies |\psi\rangle\in V_{g_1}\wedge|\psi\rangle\in V_{g_2}\wedge\cdots\wedge|\psi\rangle\in V_{g_l}\implies |\psi\rangle\in\bigcap_{g_i\in S}V_{g_i}
\end{split}  
\end{equation}
where $\wedge$ means logical conjunction. So
\begin{equation}
    V_S\subset\bigcap_{g_i\in S}V_{g_i}
\end{equation}
From set theory:
\begin{equation}
    (A\subset B)\wedge (B\subset A)\iff A=B
\end{equation}
Then \begin{equation}
     V_S = \bigcap_{g_i\in S}V_{g_i}
\end{equation}
\end{proof}

The operator
\begin{equation*}
    P_S = \frac{1}{|S|}\sum_{g\in S} g
\end{equation*}
connects stabilizer $S$ and corresponding $V_S$, because $\mathrm{Range}(P_S) = V_S$\cite{Caves}. 

\begin{property}
    $P_S^2 = P_S$
\end{property}
\begin{proof}
    \begin{align}
      P_S^2 & = \frac{1}{|S|^2}(\sum_{i=1}^k g_i)(\sum_{j=1}^k g_j)\\
    & = \frac{1}{|S|^2}(\underbrace{g_1(\sum_{j=1}^k g_j)}+\underbrace{g_2(\sum_{j=1}^k g_j)}+\cdots+\underbrace{g_k(\sum_{j=1}^k g_j)})
\end{align}
    Using the rearrangement theorem in group theory:
    \begin{equation}
        g\cdot S = S,\forall g\in S
    \end{equation}
That is, left or right multiplication by any fixed group element induces a permutation of the underlying set of the group.
\begin{equation}
    P_S^2 = \frac{1}{|S|^2}\cdot|S|\cdot(\sum_{j=1}^k g_j) = \frac{1}{|S|}\sum_{j=1}^k g_j = P_S
\end{equation}
\end{proof}

And from linear algebra, if $P^2 = P$, we have $\mathrm{Tr}P = \dim
 \mathrm{Range}(P)$\cite{Axler}, recall that $\mathrm{Range}(P_S) = V_S$, so we have $\mathrm{Tr}P_S = \dim(V_S)$.\\
 \section{Trivial and Nontrivial stabilizer}
 Not every stabilizer interest us, because some of them correspond zero vector space ($\dim V_S = 0$). We call $S$ is \textbf{trivial stabilizer} if $\dim V_S = 0$, otherwise, we call them \textbf{nontrivial stabilizer}.\\
 \begin{proposition}
     Nontrivial stabilizer $\iff \alpha I \notin S$(all $ \alpha \neq 1$) 
 \end{proposition}
 \begin{proof}
     we can easily see that $S$ is nontrivial stabilizer$\implies$ $\alpha I\notin S$(all $\alpha \neq 1$). 
 \\
 \vspace{0.5em}
 In another direction, recall that $\mathrm{Tr}P_S = \dim(V_S)$, so if stabilizer $S$ satisfies $\alpha I\notin S$(all $\alpha \neq 1$), we have
\begin{equation*}
    \dim V_S = \mathrm{Tr}P_S = \frac{\mathrm{Tr}(\sum_{g\in S} g)}{|S|} = \frac{k_1k_2\cdots k_n}{|S|}
\end{equation*}
The reason why the last equality is true is that the only operator whose trace is not 0 in $S$ is $I$ and $\mathrm{Tr}(I) = k_1k_2\cdots k_n$ .
\\
\vspace{0.5em}
Because $\dim V_S = \frac{k_1k_2\cdots k_n}{|S|} \neq 0$, so $\alpha I\notin S$ (all $\alpha \neq 1$) $\implies$ $S$ is nontrivial stabilizer.\\

In conclusion, $S$ is nontrivial stabilizer $\iff$ $\alpha I\notin S$(all $\alpha \neq 1$).
\end{proof}
Remarks: \begin{itemize} 
    \item Notice that if S is nontrivial stabilizer, then $S$ must be abelian.
    \item For $G_{2^n}$ case, we can simplify the above condition to $S $ is nontrivial
$\iff$ $ -I\notin S$, because $-I\notin S \implies \pm iI\notin S$.
\end{itemize}

\
From the above proof, Nontrivial stabilizer $\iff \alpha I \notin S$(all  $\alpha \neq 1$) 
and 
if stabilizer $S$ satisfies $\alpha I\notin S$(all $\alpha \neq 1$), we have
\begin{equation*}
    \dim V_S = \mathrm{Tr}P_S = \frac{\mathrm{Tr}(\sum_{g\in S} g)}{|S|} = \frac{k_1k_2\cdots k_n}{|S|}
\end{equation*}

So we can extract
the following proposition:
\begin{proposition}
    For any $n$ qudits system $G_{k_1\times k_2\times\cdots \times k_n}$ with any possible dimension $k_1,k_2,\cdots k_n$, the stabilizer $S$ is nontrivial, then
    \begin{equation}
        \dim V_S = \frac{k_1k_2
        \cdots k_n}{|S|}
    \end{equation} 
\end{proposition}
\section{Generators and Independent Generators}
\subsubsection{Definition of generator}
\begin{definition}\label{definition5}
     The group $S$ is \textit{generated} by a subset $A=\{g_1,g_2,\cdots,g_l\}$ if $S$ is the set of all possible product and inverse of  $g_i$ in $A$, that is
\begin{equation}
    S = \{g_{\alpha_1}^{\beta_1}g_{\alpha_2}^{\beta_2}\cdots g_{\alpha_n}^{\beta_n}~|~ \alpha_i\in\{0,\cdots l\},\beta_i\in \{-1,1\}\ , n\in \mathbb{N}\}
\end{equation}
denoted as\cite{Dummit}
\begin{equation}
    S = \langle g_1,g_2,\cdots,g_l\rangle
\end{equation}
$A$ is called the \textbf{generating set} of $S$. $g_i$ is called \textbf{generator} of $S$. If $A = \emptyset$, we define $S = \{e\}$, where $e$ is group identity. ($\mathbb{N}\equiv \{1,2,3,\cdots\}$)

\end{definition}
\begin{itemize}
    \item $g_i$ can appear repeatedly in an expression, for example, $g\in S$ may have the expression
    \begin{equation}
        g = g_1 g_2 g_2 g_5^{-1} g_4 g_5
    \end{equation}
    \item The writing order of generators in $\langle g_1,g_2,\cdots,g_l\rangle$ is arbitrary. For example:
    \begin{equation}
        G_1 = \langle X,Y,Z\rangle = \langle Y,X,Z\rangle = \langle Z,Y,X\rangle
    \end{equation}
\end{itemize}
In group theory, the following simplified representation is used
\begin{equation}
    \underbrace{g\cdot g\cdots g }_{n}= g^n,g^n\cdot g^m=g^{n+m}
\end{equation}
So some adjacent identical elements can be merged, for example:
\begin{equation}
    g = g_1 g_2 g_2 g_5^{-1} g_4 g_5 = g_1 g_2^2 g_5^{-1} g_4 g_5
\end{equation}
Through this convention, the above definition can be equivalently written as
\begin{equation}
    S =\{ g_{\alpha_1}^{\beta_1}g_{\alpha_2}^{\beta_2}\cdots g_{\alpha_n}^{\beta_n}, \alpha_i\in\{0,\cdots l\},\beta_i\in \mathbb{Z},g_{\alpha_i}\neq g_{\alpha_{i+1}}\}
\end{equation}
Let us consider various special expressions of $S$ in some special groups.
First consider a \textit{finite group}. For a finite group, the \textit{order} of any group element is finite
\begin{equation}
    g^n = e,n\in \mathbb{N}
\end{equation}
which means
\begin{equation}
    g^{-1} = g^{n-1}
\end{equation}
So for finite groups $ S = \langle g_1,g_2,\cdots,g_l\rangle $
\begin{equation}
     S = \{g_{\alpha_1}^{\beta_1}g_{\alpha_2}^{\beta_2}\cdots g_{\alpha_n}^{\beta_n}, \alpha_i\in\{0,\cdots l\},\beta_i\in \mathbb{N},g_{\alpha_i}\neq g_{\alpha_{i+1}}\}
\end{equation}

If group is an \textit{abelian group}, then the position of the group elements can be changed arbitrarily, and we can sort the expressions of any element in the order in which the generator appears in the description of the group. Specifically, if $S = \langle g_1,g_2,\cdots,g_l\rangle$, then
\begin{equation}
    S = \{g_1^{\beta_1}g_2^{\beta_2}\cdots g_l^{\beta_l},\beta_i\in \mathbb{Z}\}
\end{equation}
So if the group is a \textit{finite abelian group}, combining the above two situations, for $S = \langle g_1,g_2,\cdots,g_l\rangle$
\begin{equation}\label{eq62}
     S = \{g_1^{\beta_1}g_2^{\beta_2}\cdots g_l^{\beta_l},\beta_i\in \mathbb{N}_0\}
\end{equation}
Where $\mathbb{N}_0\equiv \{0,1, 2, 3,\cdots\}$.

\subsubsection{Independent generating set}
\begin{definition}
    Suppose $S = \langle g_1,g_2,\cdots,g_l\rangle$. If deleting any generator $g_i$ leads to 
    \begin{equation}
       \langle g_1,g_2,\cdots,g_l\rangle \neq \langle g_1,g_2,\cdots,g_{i-1},g_{i+1},\cdots,g_l\rangle 
    \end{equation}
then the generating set $\{g_1,g_2,\cdots,g_l\}$ is called an \textbf{independent generating set}. Otherwise it is called \textbf{dependent generating set}.
\end{definition}
\begin{itemize}
    \item The above definition can also be equivalently expressed as-There are $g_i\in \{g_1,g_2,\cdots,g_l\}$ which can be generated by other generators.That is 
    \begin{equation}
        g_i = g_{\alpha_1}^{\beta_1}g_{\alpha_2}^{\beta_2}\cdots g_{\alpha_{i-1}}^{\beta_{i-1}}g_{\alpha_{i+1}}^{\beta_{i+1}}\cdots g_{\alpha_l}^{\beta_l} 
    \end{equation}
    for some $\alpha_i\in\{0,\cdots,i-1,i+1\cdots,l\},\beta_i\in \{-1,1\}\ , n\in \mathbb{N}$.
\end{itemize}
\section{A Conjecture}
\begin{Conjecture}
    For $G_{k_1\times k_2\cdots\times k_n}$, if there is no relation $k_1 = k_2 = k_3=\cdots =k_n$, then there is no \textit{functional relation} between number of independent generators $l$ and order of $S$, for any nontrivial stabilizer $S$.
\end{Conjecture}
In qubits system,we have $|S| = 2^l$. But for example, to $\mathbb{C}^2\otimes\mathbb{C}^3$ system, \begin{center}
    
\begin{tabular}{|c|c|c|}
    \hline
    \textbf{$S$} & 
    $l$ & $|S|$ \\
    \hline
    $\langle X\otimes X_3\rangle$ & 1 & 6 \\
    $\langle I\otimes X_3\rangle$ & 1 & 3 \\
    $\langle Z\otimes X_3\rangle$ & 1 & 6 \\
    $\langle Z\otimes I, I\otimes X_3\rangle$ & 2 & 6 \\
    \hline
\end{tabular}

\end{center}
Obviously, there is no functional relation between $l$ and $S$.\\

\section{Acknowledgments}
The author thanks the University of York's MSc in Mathematical Sciences for an engaging learning environment, excellent mathematical training, and exposure to frontier research projects. The author is very grateful to Matthew Pusey and Máté Farkas for their patient guidance and many insightful discussions. The author also thanks Markus Grassl for pointing out reference~\cite{Wang} and for valuable feedback. The author acknowledges ChatGPT for help with language polishing in parts of the text.

\end{document}